\documentclass[10pt]{article}

\usepackage{amsfonts}
\usepackage{epsfig}

\begin{document}

\title{Markov Chain Analysis of Musical Dice Games}

\vspace{1cm}

\author{J. R. Dawin${}^{1}$,
D. Volchenkov${}^{2}$
\footnote{E-Mail:{\it  volchenk@physik.uni-bielefeld.de}}
}
\maketitle
{\scriptsize
\begin{enumerate}
\item[${}^1$]{\it Department of Physics, Universit\"{a}t Bielefeld, Postfach 10 01 31, 33501 Bielefeld, Germany}
\item[${}^2$]{\it Center of Excellence Cognitive Interaction Technology, Universit\"{a}t Bielefeld, Postfach 10 01 31, 33501 Bielefeld, Germany }
\end{enumerate}
}

\begin{flushleft}
{\bf Keywords:} Markov chains, entropy and complexity, musical distance, 
first passage time.
\end{flushleft}

\begin{abstract}
We have studied 
entropy, redundancy, complexity, and first passage times 
to notes for 804 pieces of 29 composers. The successful
understanding of tonal music calls for an experienced listener, 
as entropy dominates over redundancy in musical messages.  
First passage times to notes resolve tonality and feature a composer.
We also discuss the possible distances in space of musical 
dice games and introduced the geodesic
distance based on the
 Riemann structure associated to the probability vectors
 (rows of the transition matrices).

\end{abstract}

\vspace{0.5cm}


\newpage

\section{Musical dice game as a Markov chain}
\label{sec:Introduction}
\noindent

A system for using dice 
to compose music randomly, 
without having to know neither the techniques of composition,
nor the rules of harmony, 
named {\it Musikalisches W\"{u}rfelspiel} 
({\it Musical dice game})
had become quite popular 
throughout Western Europe 
in the $18^{th}$ century \cite{Noguchi:1996}.
Depending upon the results of dice throws, 
the certain pre-composed bars of music 
were patched together 
resulting in different, 
but similar, 
musical pieces. 
 "The Ever Ready Composer of Polonaises and Minuets" 
was devised
by Ph. Kirnberger,
 as early as in 1757. 
The famous chance music machine 
 attributed to W.A. Mozart (\verb"K 516f")  
consisted of numerous two-bar fragments of music 
named after the different letters of the Latin alphabet
and destined to be combined together 
either at random, 
or following an anagram of your beloved
had been known since 1787. 

We can consider a note as 
an elementary event 
in the musical dice game, 
as notes provide 
a natural discretization 
of musical phenomena
that facilitate 
their performance and analysis.
Given 
the entire keyboard $\mathcal{K}$
of 128 notes corresponding to 
a pitch range of 10.5 octaves,
each divided into 12 semitones,
we regard a note  
as a discrete {\it random variable} $X$ 
that maps the musical event
 to a value of a $n$-set of pitches 
$\mathcal{P}=\{x_1, \ldots, x_n\}\subseteq \mathcal{K}.$
In the musical dice game,
a piece is generated by patching 
notes $X_t$ taking values from 
the set of pitches $\mathcal{P}$
that {\it sound good together}
into a  temporal  sequence $\left\{X_t\right\}_{t\geq 1}$.
Herewith, two consecutive notes, 
in which the second pitch is 
 a harmonic of the first one
 are considered to be pleasing to the ear,
and therefore can be patched
to the sequence. 
Thus tonal harmony 
sets up the {\it Markov property}
for the sequence $\left\{X_t\right\}_{t\geq 1}$
that can be assessed
in terms of 
the transition probabilities between 
consecutive notes,
in the framework of a simple  time -- homogeneous
 model called {\it Markov chain} \cite{Markov:1906},
\begin{equation}
\label{music_transition_matrix}
\begin{array}{l}
\Pr\left[ X_{t+1}\mid X_t=y,X_{t-1}=z,\ldots\right]\,=\,
\Pr\left[ X_{t+1}\mid X_t=y\right]\,=\,T_{yx}, \\
\sum_{x\in \mathcal{P}}T_{yx}=1,
\end{array}
\end{equation}
where the stochastic transition matrix
$T_{yx}$
weights the 
chance of a pitch $x$
going directly 
to another pitch 
$y$ independently of time.
The model (\ref{music_transition_matrix})
obviously 
does not impose a severe limitation on  
melodic variability, 
since there are many possible 
combinations of notes considered consonant, 
as sharing some harmonics 
and making a pleasant sound together.
The relations between notes 
in (\ref{music_transition_matrix})
are rather described in terms of 
 probabilities 
and expected numbers of random steps 
 than by physical time.
Under such circumstances,  
the actual length $N$ of a composition 
is formally put $N\to\infty,$ or
as long as you keep rolling the dice.
Markov chains are widely used in algorithmic music
 composition, as being a standard tool,
 in music mix and production software.

Interactions between 
humans via speech and music 
constitute the unifying theme of research 
 in modern communication technologies.
As with music,
 speech and written language 
also have 
the sets of rules 
(crucial for establishing effective communication)
that govern which particular combinations 
of sounds and letters 
may or may not be produced.
However, while communications
 by the spoken and written forms of human languages 
have been paid much attention  
from the very onset of
information theory \cite{Shannon:1948,Shannon:1951},
not very much is known about 
the relevant information aspects 
of music \cite{Wolfe:2002}.
Although we use 
the acoustic channel in both music and speech,
the acoustical and structural features 
we implement
 to encode and perceive 
the signals in music and speech are
dramatically different, as
"speech is communication of world view
 as the intellection of reality while 
music is communication of world view 
as the feeling of reality" \cite{Seeger:1971}.
With 
the Markov chain model
  (\ref{music_transition_matrix}), 
we could precisely quantify this difference,
since  it allows to appraise
tonal music as a generalized communication process, 
in which a composer sends a message 
transmitted by a performer to a listener.

In our work,  
 we report some results on 
the Markov chain analysis 
of
the musical dice games 
encoded by the transition matrices 
between pitches 
in the MIDI representations 

of the 804 musical compositions
attributed to 29 composers:
 J.S. Bach (371), L.V. Beethoven (58),
A.Berg (7), J. Brahms (8), D. Buxtehude (3),
 F. Chopin (26),
C. Debussy (26), G. Faur\'{e} (5), C. Franck (7),
G.F. H\"{a}ndel (45), 
F. Liszt (4), F. Mendelssohn Bartholdi (19),
C. Monteverdi (13), W.A. Mozart (51), 
 J. Pachelbel (2), S. Rachmaninoff (4),
C. Saint-Sa\"{e}ns (2),  
E. Satie (3), A. Sch\"{o}nberg (2),
 F. Schubert (55), R. Schumann (30),
A. Scriabin (7), D. Shostakovitch (12),
J. Strauss (2),  I. Stravinsky (5),
P. Tchaikovsky (5), J. Titelouze (20), 
A. Vivaldi (4), R. Wagner (8).
The MIDI representations of 
many musical pieces are freely available on the Web
\cite{Mutopia}.

The paper is organized as follows.
In Sec.~\ref{sec:How_did_we_collect_the_data},
we discuss the
MIDI representations of music
  and the different methods 
to encode them into a Markov chain transition matrix.
The encoding problem is not trivial, as
ambiguities would arise provided a piece has 
more than one voice.  
We then consider a music as a
generalized communication process 
in Sec.~\ref{sec:Entropy_redundancy_complexity}.
While the elements of 
the transition matrix (\ref{music_transition_matrix})
indicate the possibility 
to consequently 
find the two notes in 
the musical score, 
an infinite number of powers 
of the transition matrix
must be considered to estimate the eventual 
distance between them with 
respect to 
the entire structure of the musical dice game.
First passage times to notes and 
the classification of composers with respect to 
their tone scale preferences 
 are discussed in Sec.~\ref{sec:First_Passage_Times_in_music}.
The possible distances between the different 
musical dice games are discussed in 
Sec.~\ref{sec:musical_distances}.
We conclude in the last section.

\section{Encoding of a discrete model of music (MIDI) into 
a transition matrix}
\label{sec:How_did_we_collect_the_data}
\noindent

While analyzing the statistical structure of 
musical pieces,
we used the MIDI representations
providing a computer readable 
{\it discrete time 
model} of music
by a sequence of
the 'note' events,
\verb"note_on" and \verb"note_off":
\begin{table}[ht]
\centering
\begin{tabular}{l|l|l|l|l}
\hline
\verb"event type" & \verb"time"& \verb"channel" &\verb"note"&
 \verb"velocity"\\
\hline
\verb"note_on" &  \verb"192" & \verb"0" &
\verb"60" & \verb"127" \\
\verb"note_off" &  \verb"192" & \verb"0" &
\verb"60" & \verb"64" \\
\end{tabular}
\caption{\footnotesize The MIDI events for 
the note  C4.\label{Tab_music_01}}
\end{table}
In the MIDI representation, 
each event (like that one shown in Tab.~\ref{Tab_music_01}) 
is characterized by 
the four variables:  'time', 'channel',
'note', and 'velocity'.
A MIDI file
has a specific value
of discreteness 
 'ticks/quarter' 
indicating 
the number of 'ticks' that
make up a quarter note. 
The value of  'time' 
then gives the number of 'ticks' between
two consequent note events.
In the example given in Tab.~\ref{Tab_music_01},
the event of \verb"C4" starts after 192 'ticks' have passed.
 The 'channel' indicate one of 16 channels (0 to 15)
 this event may
belong to. 
Notes are not
encoded by their names like \verb"C" or \verb"A".
Instead, the harmonic scale is mapped onto
numbers from 0 to 127 with chromatic steps.
For instance, the identification number
 60 corresponds
to the \verb"C4", in musical notation. 
Then, note number 61 is \verb"C4#", 
62 is \verb"D4" etc. (see Tab.~\ref{Tab_music_02}
for some octaves and their MIDI note ID numbers)

\begin{table}[ht]
\centering
\begin{tabular}{l|l|l|l|l|l|l|l|l|l|l|l|l}
\hline
\verb"Octave" & \verb"C" &\verb"C#"& \verb"D"& \verb"D#" &\verb"E"
 &\verb"F"& \verb"F#" &\verb"G" & \verb"G#" &\verb"A"
 &\verb"A#" &\verb"B"\\
\hline
3 & 48 & 49 &50 &51 &52 &53 &54 &55 &56 &57 &58& 59 \\
4& 60& 61& 62 &63& 64& 65& 66 &67& 68& 69& 70& 71 \\
5 &72& 73& 74& 75 &76 &77& 78& 79& 80& 81& 82& 83 \\
\end{tabular}
\caption{\footnotesize MIDI note ID numbers corresponding 
to musical notation. \label{Tab_music_02}}
\end{table}
Finally, the 'velocity' (0 -- 127)
 describes the strength with which the note is played. 
As MIDI files
 contain all musically relevant
data, it is possible 
to determine the probabilities
 of getting from one note to another
for all notes in a musical composition
by analyzing its MIDI file with
a computer program.
To get transition matrices
(\ref{music_transition_matrix})
 for tonal sequences, 
we need
 only  'time', 'channel', and
'note' to be considered.

The MIDI files of 804 musical compositions 
were processed by a program written
in \verb"Perl"; the MIDI parsing was 
done using 
the module \verb"Perl::MIDI" \cite{Perl_MIDI},
  which allowed
the conversion of the MIDI data 
into a more convenient form called
\verb"MIDI::Score"
where each two consequent 
 \verb"note_on" and \verb"note_off"
 events are
combined to a single \verb"note" event. 
Each \verb"note" event contains an absolute \verb"time",
the starting time of the event, 
and a \verb"duration" which gives the duration of the
event in ticks.

\begin{figure}[ht]
 \noindent
\begin{center}
\epsfig{file=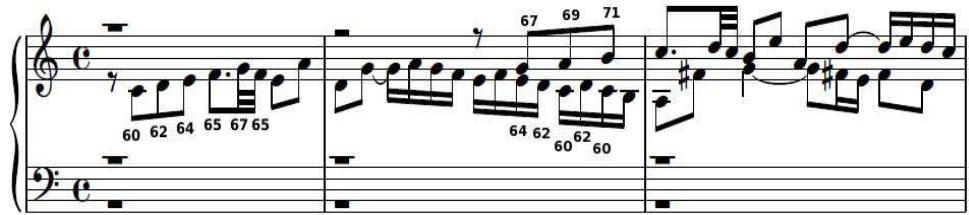,  angle= 0,width =13cm, height =3cm}
  \end{center}
\caption{\footnotesize  The first three bars from the fugue of BWV846.
 Also shown are MIDI
note numbers .}
 \label{Fig_music_02}
\end{figure}

To give an example of the process
 of getting to a transition matrix from a
musical score, we consider the first bars
of the Fugue \verb"BWV846" of J.S. Bach
shown in Fig.~\ref{Fig_music_02}.
 The numbers below the first notes
in Fig.~\ref{Fig_music_02} indicate the
corresponding
 MIDI ID note numbers.
In Tab.~\ref{Tab_music_03}, we
show the representation
 of these notes in MIDI
and in \verb"MIDI::Score" format. 
Here, the value of velocity is omitted.

\begin{table}[ht!]
\begin{tabular}{l|l|l|l||l|l|l|l|l}
 \hline
\multicolumn{4}{l|}{"MIDI"}& \multicolumn{5}{||l}{"MIDI::Score format"} \\
\hline
 \verb"event type" & \verb"time" & \verb"ch" & \verb"note"  & 
\verb"event type" & \verb"time"& \verb"dur" & \verb"ch" & \verb"note"\\
\hline
\verb"note on" &  192  & 0  & 60 & \verb"note" & 192  & 192 & 0  & 60 \\
 \verb"note off"    &   192 &   0 &   60 &     &  & & & \\
\verb"note on" &          0 &    0 & 62    &        \verb"note"    &    384  & 192& 0 &  62\\
\verb"note off" &        192 &   0 &  62&     &  & & & \\
\verb"note on" &          0  &   0 &  64     &       \verb"note"   &     576  & 192& 0 &  64 \\
\verb"note off" &         192 &  0 &  64&     &  & & & \\
\hline
\end{tabular}
\caption{\footnotesize MIDI and MIDI::Score data 
from the beginning of Fugue I BWV846.}
\label{Tab_music_03}
\end{table}

For the first notes shown in  Fig.~\ref{Fig_music_02}, 
the definition of a transition
 is easy as there is only
one voice. 
In particular, from Tab.~\ref{Tab_music_03}, 
we can conclude that  
there would be the consequent 
 transitions $60 \to 62$ and
$62 \to 64$. 
However,  
like most musical pieces, this 
Fugue then 
 contains several voices that
play simultaneously, so that 
an additional convention
is required to define a transition from note
to note.

 \begin{table}[ht]
\centering
\begin{tabular}{llll|ll}
\hline 
\verb"time" & \verb"dur" &   \verb"ch"
 &\verb"note"  &\verb"name" & \verb"voice"
\\
\hline\hline
2496  & 192 & 0& 67& \verb"G4"& upper \\
2496 & 96 & 0 & 64& \verb"E4"& lower   \\
2592 & 96& 0 & 62& \verb"D4" &lower    \\
2688& 192& 0 &69 &\verb"A4" &upper   \\
2688 & 96& 0 &60 &\verb"C4" &lower      \\
2784 &96 &0 &62 &\verb"D4"& lower        \\
2880 &192& 0 &71& \verb"B4"& upper        \\
2880& 96& 0 &60& \verb"C4"& lower      
\end{tabular}
\caption{\footnotesize  MIDI::Score data from the
 middle of the second bar of Fugue I
BWV846 where the second voice 
starts playing. The note names and the voices
of the events are also shown.
 \label{Tab_music_04}}
\end{table}
In the middle of the second bar
shown in Fig.~\ref{Fig_music_02},
 a second voice is starting. Some note
events starting from there are given 
in Tab.~\ref{Tab_music_04} in \verb"MIDI::Score" form.
From Tab.~\ref{Tab_music_04}, it is clear 
that in principle 
it is not  necessary to
put the upper voice into a different
 channel than that of the lower voice. 
In the example shown in Tab.~\ref{Tab_music_04}, 
the notes 67 and 64 both start at time 2496. 
As note 64 has a duration
of 96 ticks, it is obvious that note 
62 at time 2592 belongs to the same voice
as note 64. 
However,
 for the notes 69 and 60 starting at 2688,
 it is unclear to which
voice each note belongs to, and how
they might be encoded into 
a transition matrix.
It is important to note
that such an ambiguity 
 is not a problem of MIDI representation
itself,
 but rather
of
music.
It 
depends upon the experience of a listener
how
she distinguishes voices
while 
listening to a musical composition
 that contains several simultaneous voices.
 Even if the
musical score 
explicitly separates 
those voices 
by placing them atop of each other, 
our personal impression of them
might not coincide with that one notated,
rather arising from live audio mixing 
of all simultaneous voices 
during the performance.
Thus, to get transition matrices from 
MIDI fies, 
we have to answer the following important 
question: "Which transitions between
which note events have to 
 be accounted?"

In our approach,
 we sort
note events ascending
by time and channel.
By surfing over the list of events,
a transition between two
subsequent occurrences
is accounted
when the moment of time
 of the second event is greater
than that 
of the first one.
When several events
occur 
simultaneously, 
we give  
a priority 
to the event belonging to 
a small number
channel. 
Let us emphasize that  
under the used method 
not all possible transitions
between note events
 contribute into the transition matrix.
\begin{figure}[ht!]
  \begin{minipage}{.5\linewidth}
      \centering
\epsfig{file=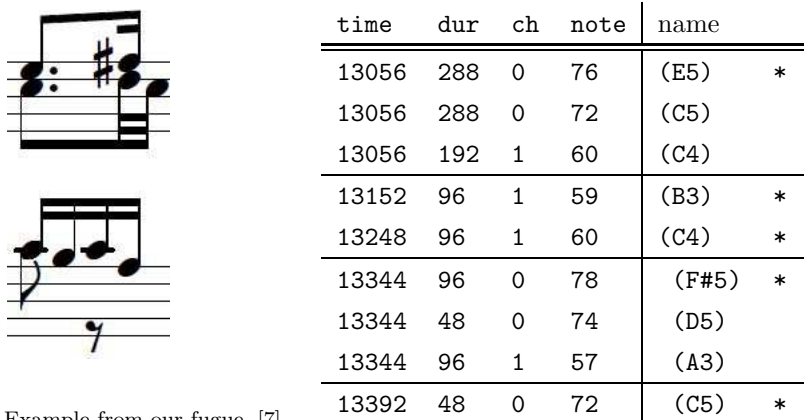, scale=0.5, angle=0}    
\hfill
      \caption{\footnotesize Example from our fugue, \cite{Mutopia}. 
\label{Fig_music_02a}}
  \end{minipage}%
  \begin{minipage}{.5\linewidth}
    \tt
    \begin{tabular}{llll|ll}
      
      time & dur & ch & note & {\rm name}& \\
      \hline\hline
      13056 & 288 & 0 & 76 & {(\verb"E5")} & * \\
      13056 & 288 & 0 & 72 & {(\verb"C5")} \\
      13056 & 192 & 1 & 60 & {(\verb"C4")} \\
      \hline
      13152 & 96 & 1 & 59 & {(\verb"B3")} & *\\
      13248 & 96 & 1 & 60 & {(\verb"C4")} & *\\
      \hline
      13344 & 96 & 0 & 78 & { (\verb"F#5")} & *\\
      13344 & 48 & 0 & 74 & { (\verb"D5")} \\
      13344 & 96 & 1 & 57 & { (\verb"A3")} \\
      \hline
      13392 & 48 & 0 & 72 & { (\verb"C5")} & *\\
    \end{tabular}
    \rm
\label{Tab_music_05}
  \end{minipage}
\end{figure}
For example, let us consider
 the musical bar shown
 in Fig.~\ref{Fig_music_02a}
 and its list of
events given in the adjacent table.
The resulting transitions 
accounted in the matrix 
would be those between events marked with  '*':
$76 \to 59$, $59 \to 60$, $60 \to 78$, $ 78 \to 72.$
Note events with small
channel values are favored over 
those with higher values. 
For simultaneous
note events occurring 
in the same channel, 
only the first one 
is considered 
that mostly means the topmost voice,
 in
musical notation.  
\begin{figure}[ht!]
 \noindent
\begin{center}
\begin{tabular}{lr}
\epsfig{file=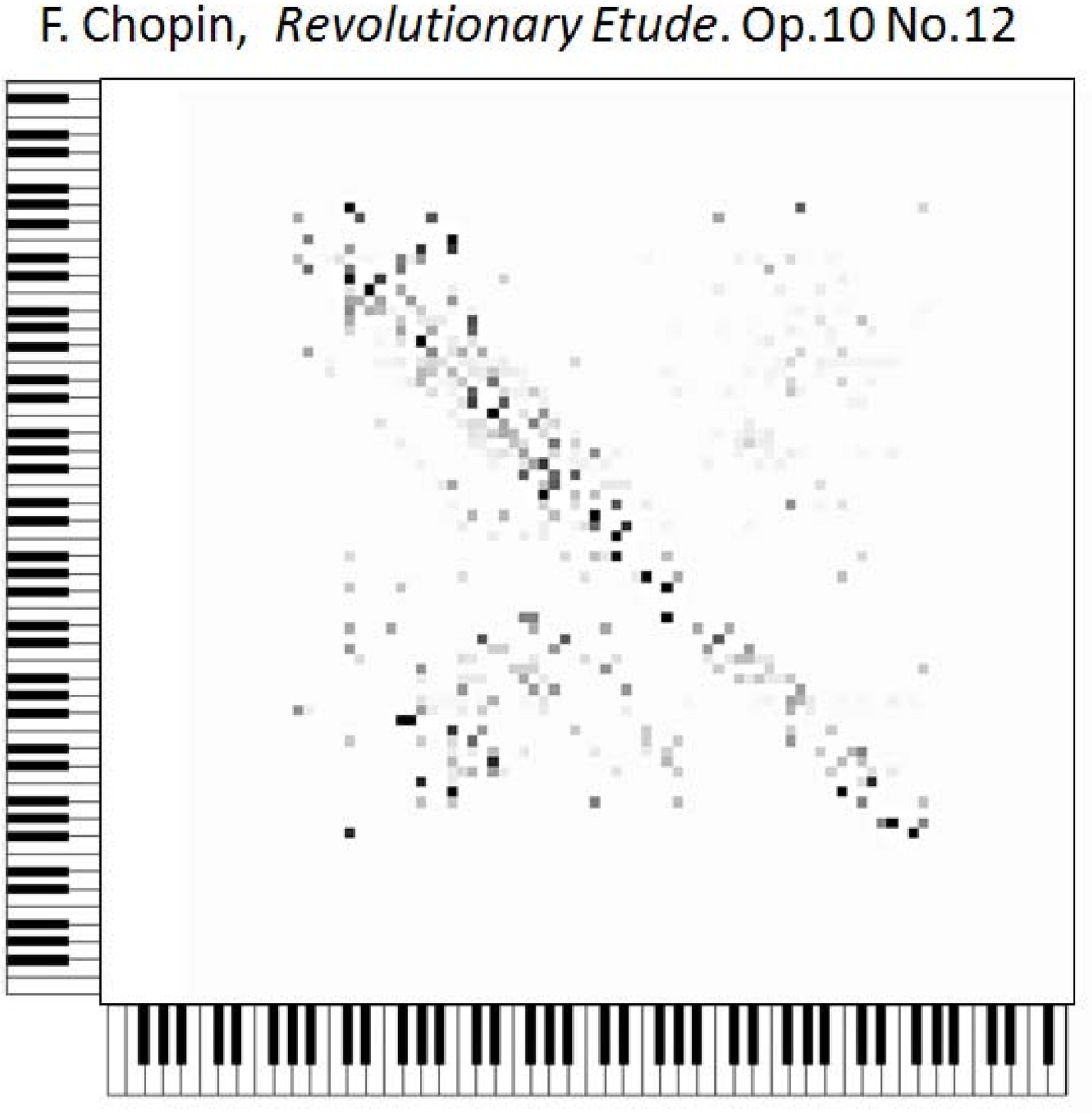, width=6cm, height =6cm}& 
 \epsfig{file=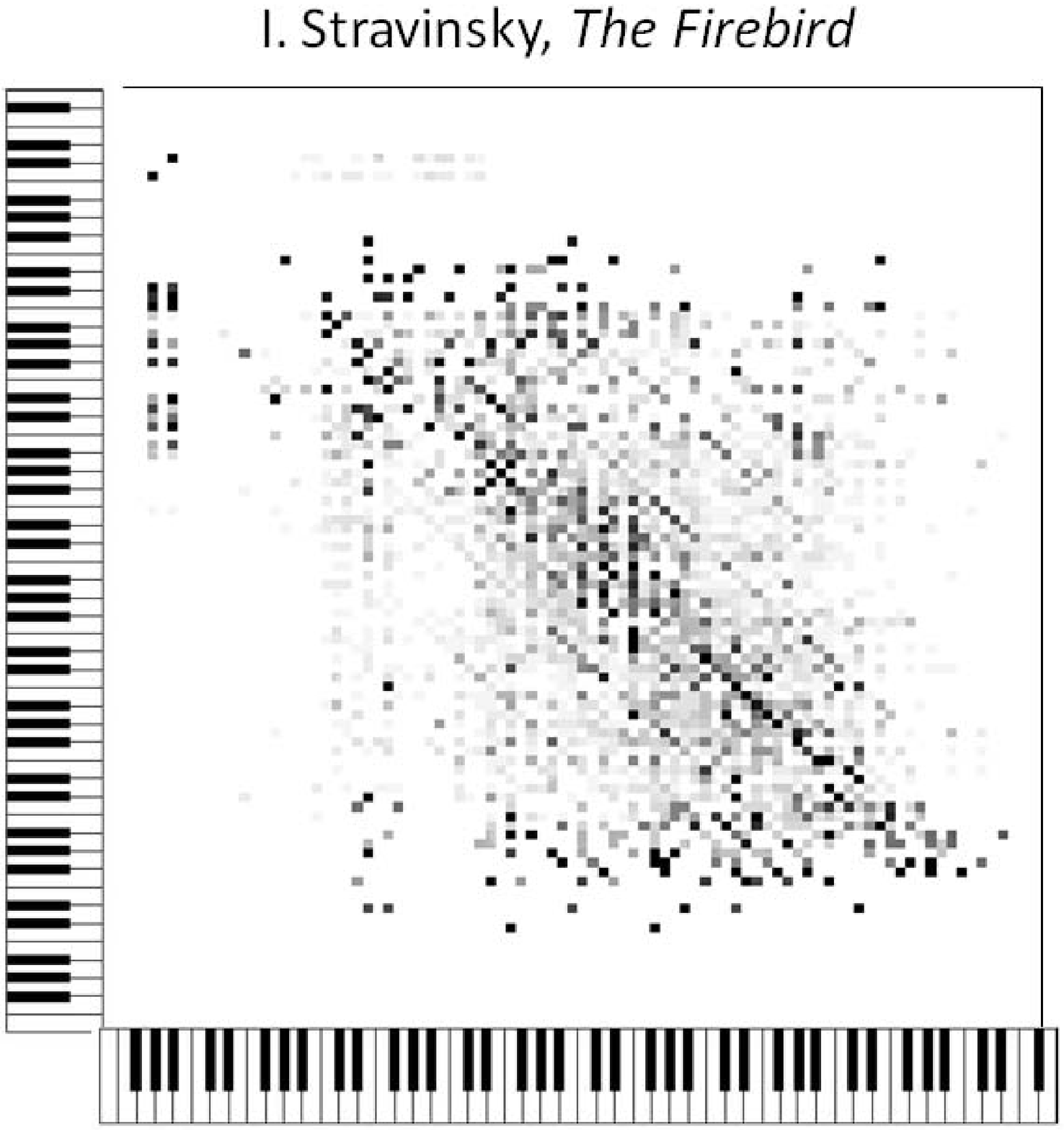, width=6cm, height =6cm} 
\end{tabular}
\end{center}
\caption{\footnotesize Transition matrices  
for the F.Chopin "Revolutionary Etude" (Op.10, No 12) (left)
and the I. Stravinsky "The Fire-bird" suite (right) generated 
accordingly to the encoding method we used.
 \label{Fig_music_03}}
\end{figure}
We believe that 
the encoding method we use 
is quite efficient for unveiling the 
individual melodic lines and identifying a
creative character of a composer from 
musical compositions 
because of the appearance of the resulting 
    transition matrices.
Those matrices 
generated 
with respect to 
the chosen encoding method
   look differently, from piece to piece and 
from composer to composer 
 (see the examples shown in Fig.~\ref{Fig_music_03}). 
However, if we were treated
each voice in a musical composition 
separately
(the transitions of the upper voice
and those
 of the lower voice might be accounted independently
while computing the probabilistic vector 
forming a row 
of the transition matrix),
the transition matrices were 
clearly dominated by a region along the main diagonal,
similarly for all compositions.

It is important to mention that 
no mater which encoding method is used 
the resulting transition matrices 
appear to be essentially not symmetric:
if $T_{xy}>0,$ for some $x,y,$
 it might be that $T_{yx}=0.$
A musical composition can be represented 
by a weighted directed graph, in which 
vertices are associated with pitches 
and directed edges connecting them 
are weighted accordingly to the 
probabilities of the immediate transitions 
between those pitches.
  Markov's chains
determining random walks 
on such the graphs are not ergodic: 
 it may
  be impossible to go
 from every
 note  
to every other note 
following the score of the musical piece.

\section{Musical dice game as a communication process}
\label{sec:Entropy_redundancy_complexity}

\begin{figure}[ht!]
 \noindent
\begin{center}
\epsfig{file=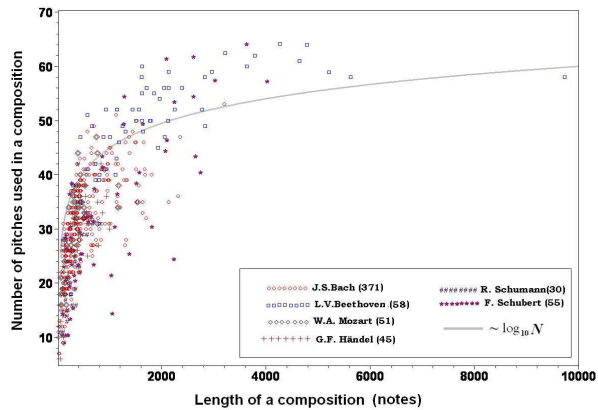,  angle= 0,width =8cm, height =5.5cm}
  \end{center}
\caption{\footnotesize The number of different pitches
  used in a composition 
grows approximately logarithmically with the size of compositions.
 The data have been collected over 371 chorales of J.S. Bach, 
58 various compositions of L.V. Beethoven, 
55 compositions of F. Schubert,
51 compositions of W.A. Mozart, 
45 compositions of G.F. H\"{a}ndel, 
and 30 compositions of R. Schumann.}
 \label{Fig_music_01}
\end{figure}

Contrary to the alphabets
used in human languages,
the sets of pitches
underlying the different 
musical compositions can 
be very distinct and
may not overlap 
(even under chromatic transposition). 
The cardinality of the set 
of pitches $\mathcal{P}$
also changes from piece to piece
demonstrating a tendency of slow growth, 
with the length of composition $N$. 
In Fig.~\ref{Fig_music_01},
we have sketched how 
the number of different pitches
$n$ 
used to compose a piece 
depends upon the size of composition $N$.
The data collected over 610 pieces 
created by the six
 classical music composers
show that the growth 
can be well approximated by a
logarithmic curve indicating that 
$n\,(\sim \log N)$
can be 
used as the simplest parameter 
assessing complexity of 
a classical musical composition.
Let us suppose that 
a musical piece
 generated as an output of the musical dice game
(\ref{music_transition_matrix})
can be encoded by a sequence of 
independent and identically-distributed random variables
representing notes 
which can take values of different pitches. 
To measure the uncertainty associated with a pitch,
we can then use the Shannon entropy \cite{Schurmann:1996},
\begin{equation}
\label{musical_entropy}
H\,\,=\,\,-\sum_{x\in\mathcal{P}}
\pi_{x}\log_n\pi_{x}
\end{equation}
where $\pi_{x}$ is 
the probability to find
 the note $x\in \mathcal{P}$ 
in the musical score, 
and the base of the logarithm is 
$n =|\mathcal{P}|$.
Since the entropy of a musical 
piece defined by (\ref{musical_entropy})
 is affected by the number of used pitches
$n$, the parameter of information redundancy,
\begin{equation}
\label{music_redundancy}
R\,\,=\,\,1- {H}/{\max H}, \quad  \max H\, =\,\log n,
\end{equation}
where $ \max H$ is
 the theoretical maximum entropy,
might be used
 for comparing 
different musical compositions.
Accordingly to information theory \cite{Cover:1991}, 
redundancy quantifies predictability
of a pitch in the piece, 
as being a natural counterpart 
of entropy.  

As we have mentioned above, 
a Markov chain encoding 
the  musical dice game
is not ergodic, 
and therefore 
the probability to find a pitch in the 
musical score cannot be found simply as 
the entry in the left eigenvector of the 
transition matrix ${\bf T}$ 
belonging to the maximal eigenvalue 
$\mu=1.$
In order to find the 
probability of observing the note in the 
musical score,
we can use the  method of group generalized 
inverse \cite{Meyer:1975,Meyer:1982} that
might be applied for analyzing every Markov chain 
regardless of its structure.
As the Laplace operator
corresponding to the Markov chain 
(\ref{music_transition_matrix}),
\begin{equation}
\label{music_laplace_opeartor}
{\bf L}\,\,=\,\, {\bf 1}-{\bf T},
\end{equation}
where ${\bf 1}$ is a unit matrix, 
 is always 
a member of a multiplicative matrix group,
it always possesses a group inverse ${\bf L}^\sharp,$
a special case of the Drazin generalized inverse 
\cite{Drazin:1958,Ben-Israel:2003,Meyer:1975}
satisfying the  Erd\'{e}lyi
conditions \cite{Erdelyi:1967}:
\begin{equation}
\label{Drazin_inverse}
 {\bf LL}^\sharp {\bf L}\,\,=\,\,{\bf L}, \quad
   {\bf L}^\sharp {\bf LL}^\sharp\,\,=\,\,{\bf L}^\sharp, 
\quad 
 \left[{\bf L},{\bf L}^\sharp\right]\,\,=\,\,0
\end{equation} 
where $[{\bf A},{\bf B}]={\bf AB}-{\bf BA}$ denotes
 the commutator of 
the two matrices.
The role of group inverses (\ref{Drazin_inverse})
 in the analysis of Markov chains  
have been discussed in details in
\cite{Meyer:1975,  Meyer:1982,Campbell:1979}.
Here, we only mention that 
the stationary vector of 
a Markov chain can be calculated as
\begin{equation}
\label{music_stationary_vector}
\pi_{x_i}\,\,=\,\,\left({\bf 1}-{\bf LL}^\sharp\right)_{x_ix_j};
\end{equation}
the rows of (\ref{music_stationary_vector}) 
are all equal to the corresponding components of the 
stationary vector $\pi$. 

Determining the entropy of texts
written in a natural language
  is an important problem 
of language processing. 
The entropy of 
 current written and spoken
languages (English, Spanish) 
has been estimated experimentally 
as ranged from 0.5 to 1.3 bit per character
\cite{Shannon:1951,Lin:1973}.
An approximately even balance (50:50)
 of entropy and redundancy
is supposed 
as necessary to achieve effective 
communication in transmitting a message, 
as it makes easier for humans 
to perceive information \cite{Lin:1973}.

\begin{figure}[ht!]
 \noindent
\begin{center}
\epsfig{file=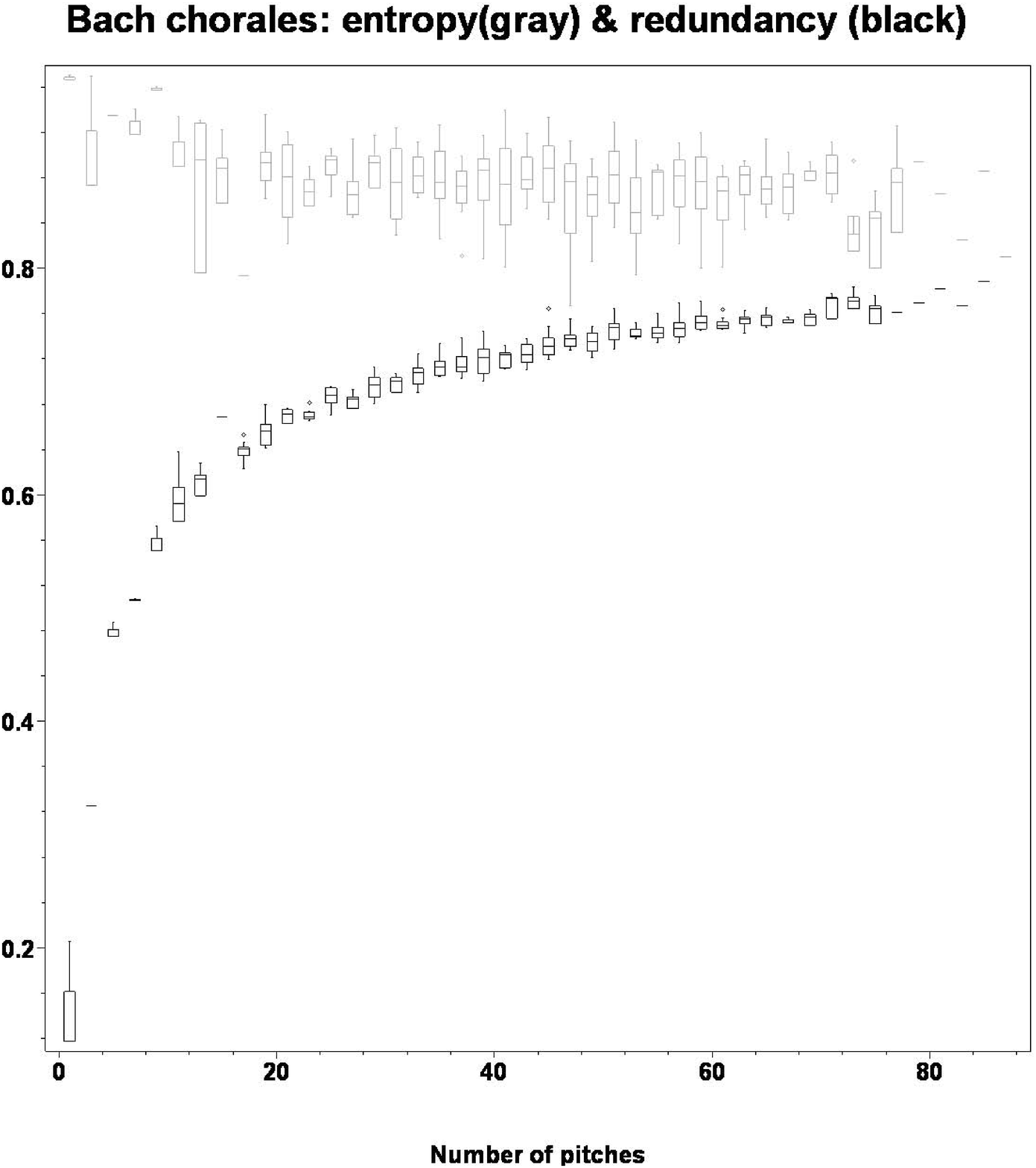,  angle= 0,width =8cm, height =8cm}
  \end{center}
\caption{\footnotesize  The box plots  
show the statistic 
of  
the magnitudes of  entropy 
(\ref{musical_entropy})
and redundancy (\ref{music_redundancy})
  vs. 
the number of pitches used in a composition,
 for 
 371 chorales of J.S. Bach.
In a box plot,  a central line of each box 
shows the median; a lower line 
shows the first quartile; 
an upper line shows the third quartile;  
 two lines extending from the central box 
of maximal length 3/2 the interquartile range 
but not extending past the range of the data; 
eventually, the outliers 
are those points lying outside 
the extent of the previous elements. \label{Fig_music_04}}
\end{figure}

For all musical compositions we studied,
the magnitudes of entropy 
fluctuate in a range 
between 0.7 and 1.1 bit per note well
fitting  with the entropy range
of usual languages. 
In classical music where the
 tonal method of composition is widely used, 
pieces involving more pitches appear to have
 lower magnitudes of entropy but higher
 values of redundancy (predictability).
In Fig.~\ref{Fig_music_04}, 
we have
presented the 
statistics
of 
entropy and redundancy 
vs. the number of pitches
through their five-number summaries,
for 371 chorales of J.S. Bach.
A central line of each box 
 in the box plot (Fig.~\ref{Fig_music_04})
 shows the median (not the mean),
the value separating the higher half of the data sample
 from the lower half, 
that is 
found by arranging all the observations from lowest value
to highest value and picking the middle one.
Other lines of the box plot 
indicate the quartile values
which divide the sorted data 
set into four equal parts, 
so that each part represents 
one fourth of the sample. 
 A lower line in each box
shows the first quartile, and 
an upper line shows the third quartile.
  Two lines extending from the central box 
of maximal length 3/2 the interquartile range 
but not extending past the range of the data. 
The outliers 
are those points lying outside 
the extent of the previous elements.

The entropy and redundancy statistics
 suggests
 that compositions in classical music
 might contain some repeated patterns,
 or motives in which certain combinations
 of notes are more likely to occur
 than others.
In particular, 
the dramatic increase of redundancy 
as the range of pitches expands up to 
7.5 octaves 
implies that 
musical compositions
involving many pitches 
might convey mostly
 conventional, 
 predictable blocks of information
to a listener.
However, 
in contrast to human languages
where  entropy and 
redundancy are 
approximately equally 
balanced \cite{Lin:1973},
in classical music
entropy clearly dominates over redundancy.
While 
decoding a musical message
requires 
the listener to invest
 nearly  
as much efforts as in
everyday decoding of speech,
 the successful understanding of
the composition would 
call for 
 an experienced listener 
ready to invest his or her 
 full attention
to a
 communication process 
that would span across cultures and epochs.

\begin{figure}[ht!]
 \noindent
\begin{center}
\begin{tabular}{lr}
\epsfig{file=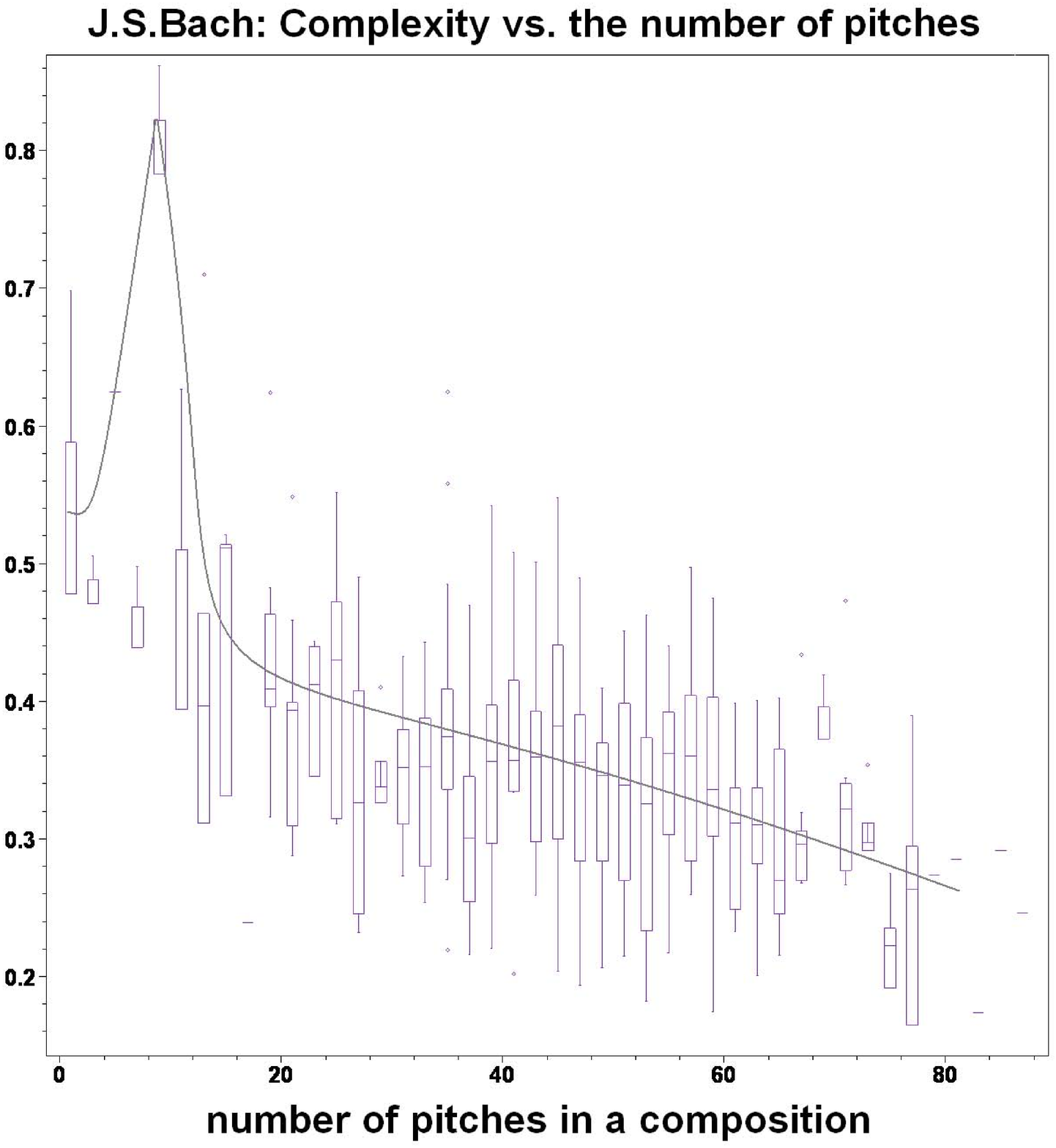, width=5.5cm, height =6cm}& 
 \epsfig{file=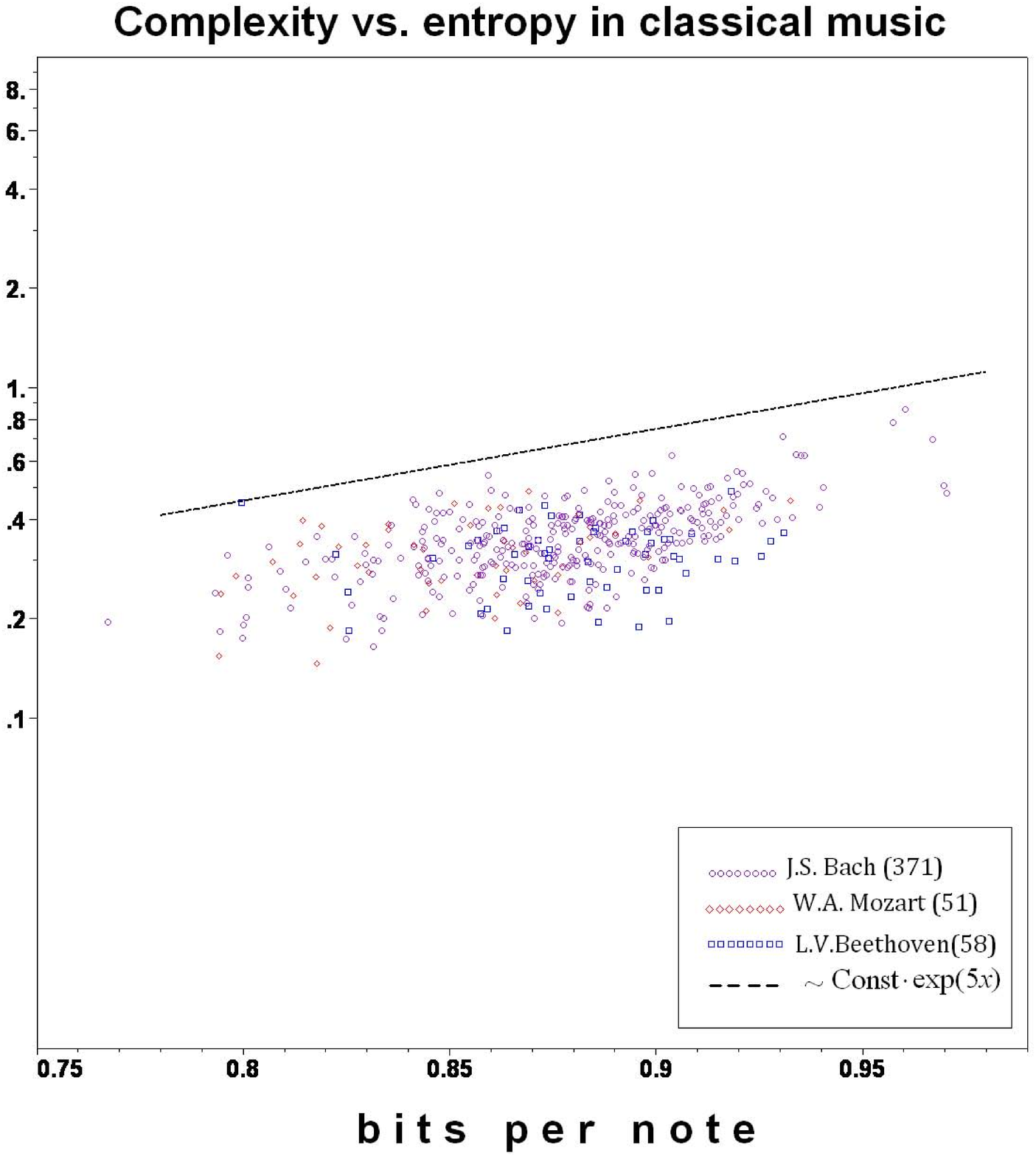, width=5.5cm, height =6cm} 
\end{tabular}
\end{center}
\caption{\footnotesize The box plot (left)
  represents  
 complexity (measured by the past-future mutual 
information) 
 vs. 
the number of pitches used in a composition,
 for 
 371 chorales of J.S. Bach.
 The trend (shown by 
the solid lines) is the cubic splines 
interpolating between the mean values 
of complexity over the data ranges. 
The scatter  plot of complexity 
vs. the magnitude of entropy in 480 pieces 
written by classical composers (given in the log-linear scale)
suggests that a strong positive correlation 
exists between  the value of
 entropy and the logarithm of complexity. 
The reference line indicates an 
exponential growth, in the log-linear scale.
 \label{Fig_music_05}}
\end{figure}
Another possible information measure 
that can be applied to 
the analysis of musical dice games
is the past-future mutual information
({\it complexity})
introduced in studies of 
the symbolic sequences generated
by dynamical systems \cite{Shaw:1984}
(see also \cite{Cover:1991}).
It estimates the information content 
of the blocks of notes and
can be formally derived as
the limiting excess
of the block entropy
$$
H(S^N)\,\,=\,\,-\sum_{S^N}P(S^N)\log_n P(S^N),
$$ 
in which $P(S^N)$ is the probability to 
find a patch $S^N$ of $N$ notes,
over the $N$ times Shannon entropy $H$,
 as the size of the block $N\to\infty,$
\begin{equation}
\label{musical_complexity_past_future}
C\,\,=\,\,\lim_{N\to \infty}
\left( H(S^N)-H\cdot N\right).
\end{equation}
Following \cite{Li:1991}, 
we use the fact that the transition 
probability between states 
in a Markov chain determined by the matrix 
(\ref{music_transition_matrix})
is independent of $N$,
so that complexity (\ref{musical_complexity_past_future}) 
can be computed simply as 
\begin{equation}
\label{musical_complexity_past_future_02}
C\,\,=\,\,-\sum_{x\in \mathcal{P}}\pi_{x}\,\log_n
\frac{\pi_x}{\,\, \prod_{y\in\mathcal{P}}T_{xy}^{T_{xy}}\,\,}.
\end{equation}
In Fig.~\ref{Fig_music_05} (left),
we have presented the statistics 
 of complexity values for the Bach's chorales.
 The main trend 
(shown in Fig.~\ref{Fig_music_05} (left) by a solid line)
is given by a cubic spline 
interpolating between the mean 
(not the median) complexity values.
Complexity
decreases 
rapidly 
with the number of pitches used in a composition 
suggesting that the musical pieces might
 contain a few types of melodic lines 
translated
over the entire diapason of pitches
by 
 chromatic transposition.
Finally, in Fig.~\ref{Fig_music_05} (right),
we have sketched a scatter plot
showing  the pace of complexity with entropy
in 480 compositions of classical music
that implies  
that a strong 
 positive correlation exists
between the value of entropy and the logarithm of complexity,
in compositions of classical music.

\section{First passage times to notes resolve tonality and
 feature a composer}
\label{sec:First_Passage_Times_in_music}

Statistics of entropy, 
redundancy, and complexity
in the discrete time models 
of  classical musical compositions  
suggests that tonal music
generated by the musical dice game
(\ref{music_transition_matrix})
 constitutes the well structured data 
that contain conventional 
patterns of information.
Obviously, 
some notes 
might be more "important" than others,
with respect to such a structure.

In music theory \cite{Thomson:1999},  
the hierarchical pitch relationships 
are introduced  based on a {\it tonic} key,
a pitch which is the lowest degree of a scale
and
that all other notes 
in a musical composition 
gravitated toward.
A successful tonal piece of music 
 gives a listener 
a feeling that 
a particular (tonic) chord 
 is the most stable and final.
The regular  method to establish
a tonic 
through a cadence, 
a succession of several 
chords which
 ends a musical section 
giving a feeling 
of closure,  
may be difficult to apply 
without listening to the piece.

\begin{figure}[ht!]
 \noindent
\begin{center}
\begin{tabular}{lr}
\epsfig{file=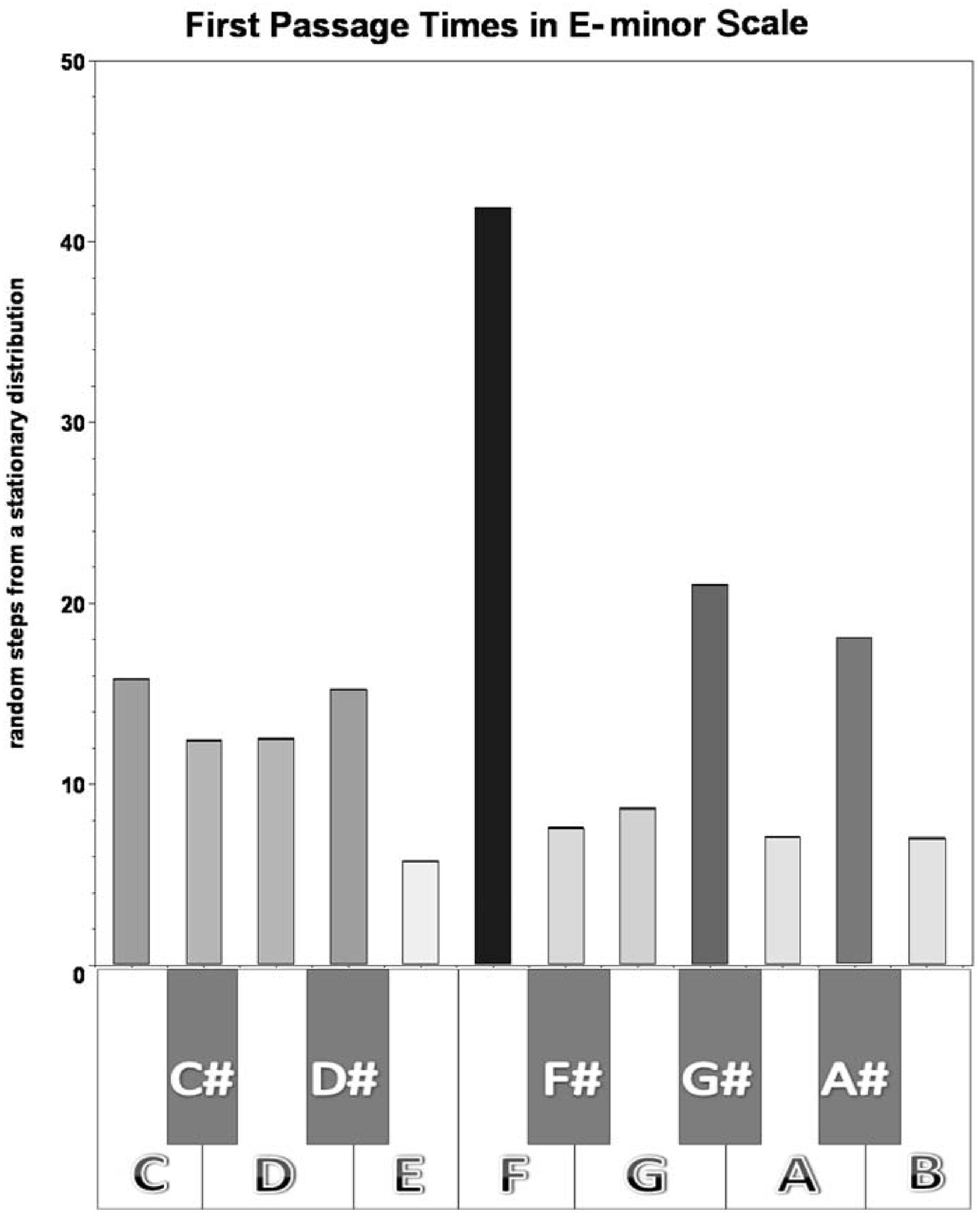, width=5.5cm, height =6cm}& 
 \epsfig{file=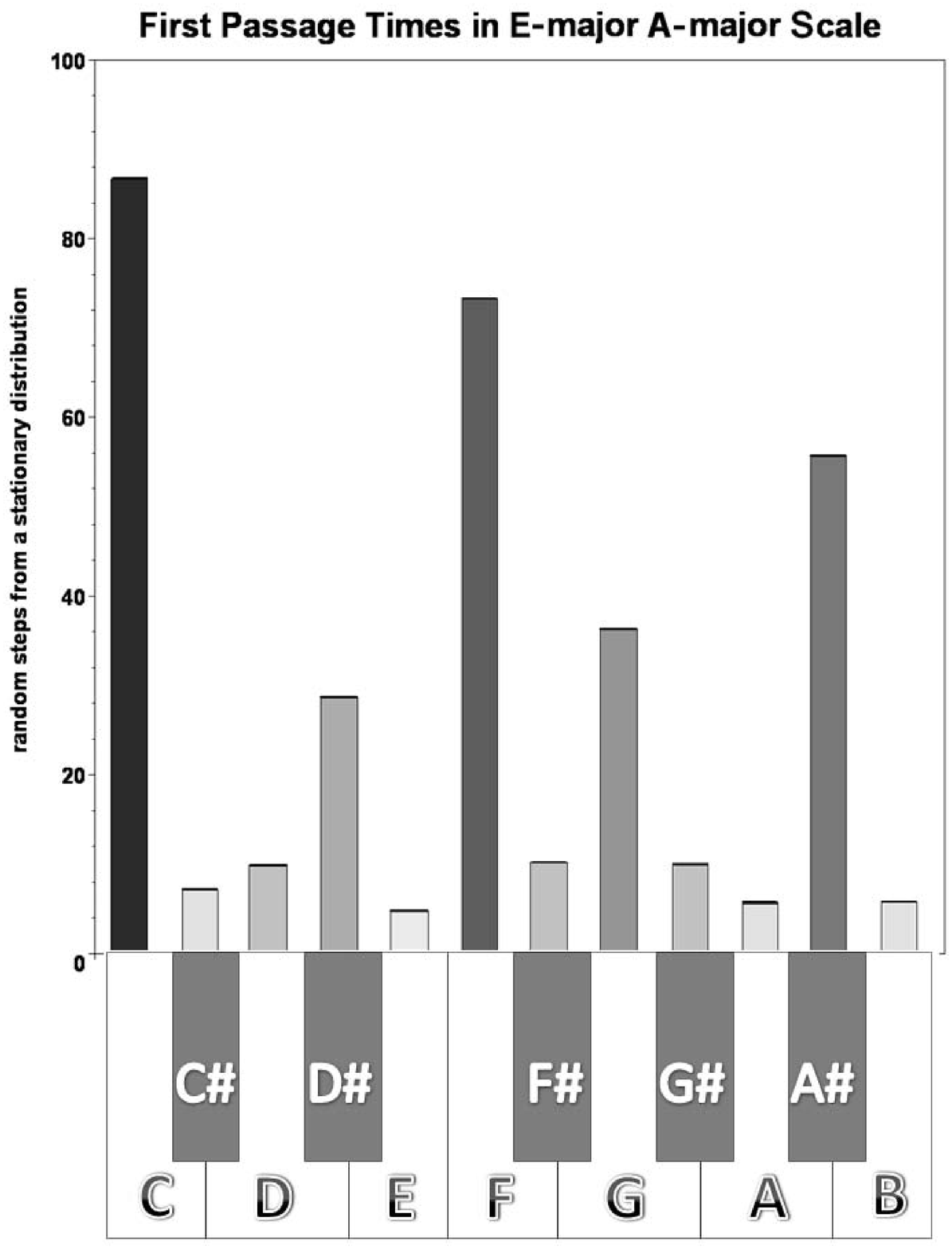, width=5.5cm, height =6cm}
\end{tabular}
\end{center}
\caption{\footnotesize 
The histograms show the first passage times to the
notes for the Duet I, BWV 802(E minor) of
J.S. Bach (left) and for the 
Cello Sonata No.3, Op.69 of 
L.V. Beethoven (E major, A major) (right)   
mapped into a single octave.
Bars are shaded
with
the intensity of gray scale 0–-100\%, 
in proportion to the magnitude of the first passage time.
Therefore, the basic pitches of a tonal scale 
are rendered with light gray color, as being 
characterized by short first passage times,
 and the tonic key by the smallest magnitude of all. 
 \label{Fig_music_06a}}
\end{figure}

While in a musical dice game, 
the intuitive vision of musicians
describing the tonic triad  
as  the "center of gravity" 
to which other chords are to lead
acquires a quantitative expression.
Namely, 
every pitch in a musical piece 
is characterized 
with respect to the entire structure of the Markov chain
by its level of accessibility 
estimated by the first passage time to it
\cite{Blanchard:2008,Volchenkov:2010},
that is the average length 
of the shortest random path 
toward the pitch from any other one
randomly chosen in the musical score.
Analyzing the first passage times 
 in scores of tonal musical compositions,
we have found that they 
can help in resolving 
 tonality of a piece, 
as they precisely render
the hierarchical relationships between 
pitches.

The majority of tonal music assumes that notes spaced 
over several octaves are perceived the same way 
as if they were played in one octave \cite{Burns:1999}. 
Using this assumption of 
octave equivalency,
we can 
 chromatically transpose
each musical piece into a single octave
getting  
the $12\times 12$ transition matrices,
uniformly for all musical pieces, independently 
of the actual number of pitches used in composition.
Given a stochastic matrix ${\bf T}$
describing transitions 
between notes within a single octave $\mathcal{O}$, 
the first passage time to the note $i\in\mathcal{O}$
is computed \cite{Volchenkov:2010} 
as the ratio of the diagonal elements, 
\begin{equation}
\label{music_first_passage_time}
\mathcal{F}_i\,\,=\,\,\left({\bf L}^{\#}\right)_{ii}/
\left({\bf 1}-{\bf LL}^{\#}\right)_{ii},
\end{equation}  
where ${\bf L}$ is the
Laplace operator corresponding to the transition matrix ${\bf T}$,
 and ${\bf L}^{\#}$ is its group generalized inverse. 
Let us note that 
in the case of ergodic Markov chains
the result (\ref{music_first_passage_time})
coincides with the classical one
on the first passage times
of random walks
defined on undirected graphs
  \cite{Lovasz:1993}.

In Fig.~\ref{Fig_music_06a}, 
we have shown the two examples of 
the arrangements of first passage times 
to notes
in one octave, for the  
 E minor scale (left) and 
E major, A major scales (right).
The 
basic pitches for the 
E minor scale 
are 
\verb"E", \verb"F#", \verb"G", \verb"A",
\verb"B", \verb"C", and \verb"D".
The E major scale is based on \verb"E", \verb"F#",
\verb"G#", \verb"A", \verb"B",
\verb"C#",  and \verb"D#". 
Finally, the A major scale consists of
\verb"A", \verb"B", \verb"C#", \verb"D",
\verb"E", \verb"F#", and  \verb"G#".
The
values of first passage times 
 are strictly 
ordered in accordance to their role in 
the tone scale of the musical composition. 
Herewith, 
the tonic key is characterized 
by the shortest first passage time 
(usually ranged from 5 to 7 random steps),
and the values of first passage times 
to other notes collected in ascending 
order reveal the entire hierarchy of their 
relationships in the musical scale.

\begin{figure}[ht!]
 \noindent
\begin{center}
\begin{tabular}{lr}
\epsfig{file=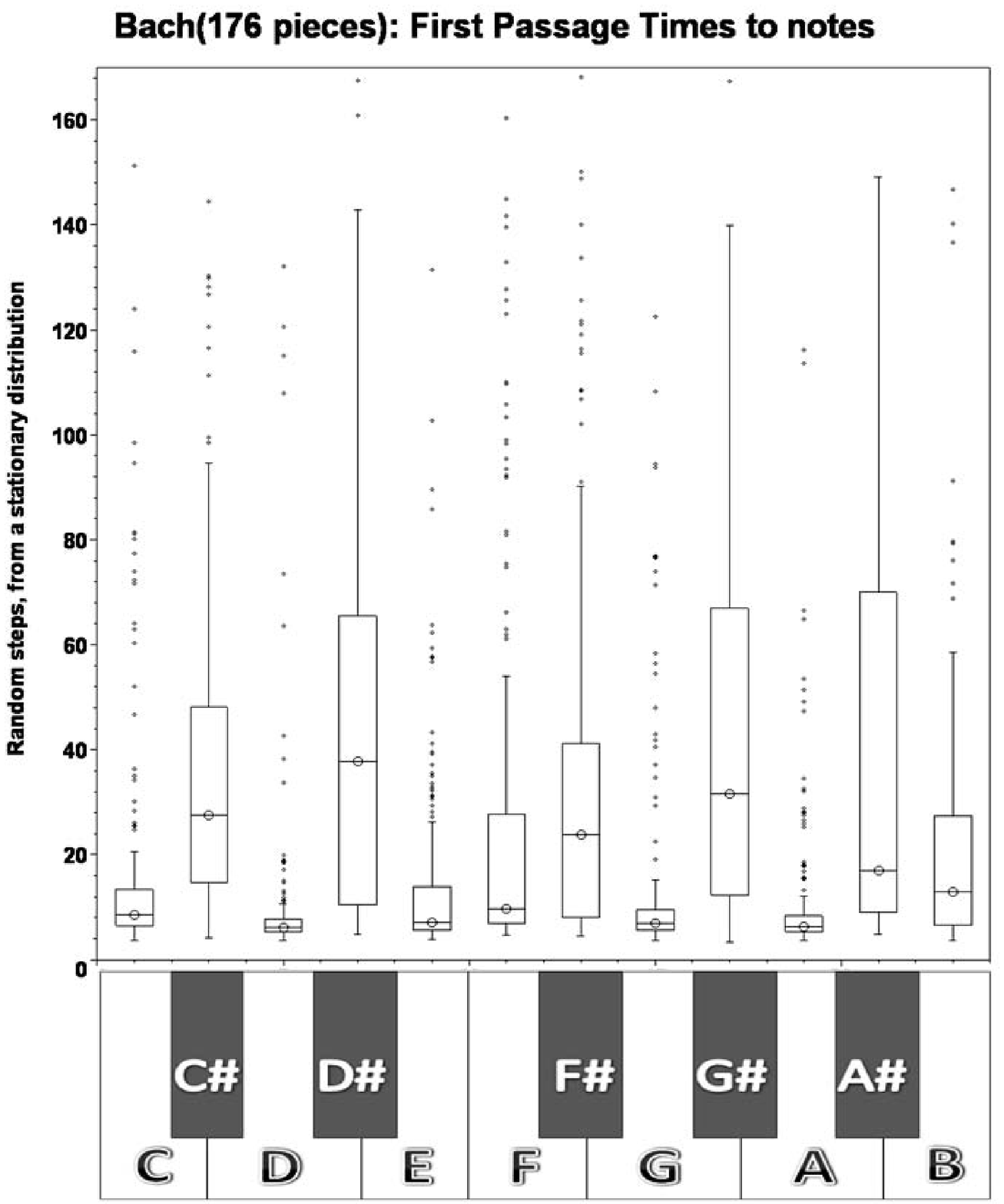, width=5cm, height =6cm}& 
 \epsfig{file=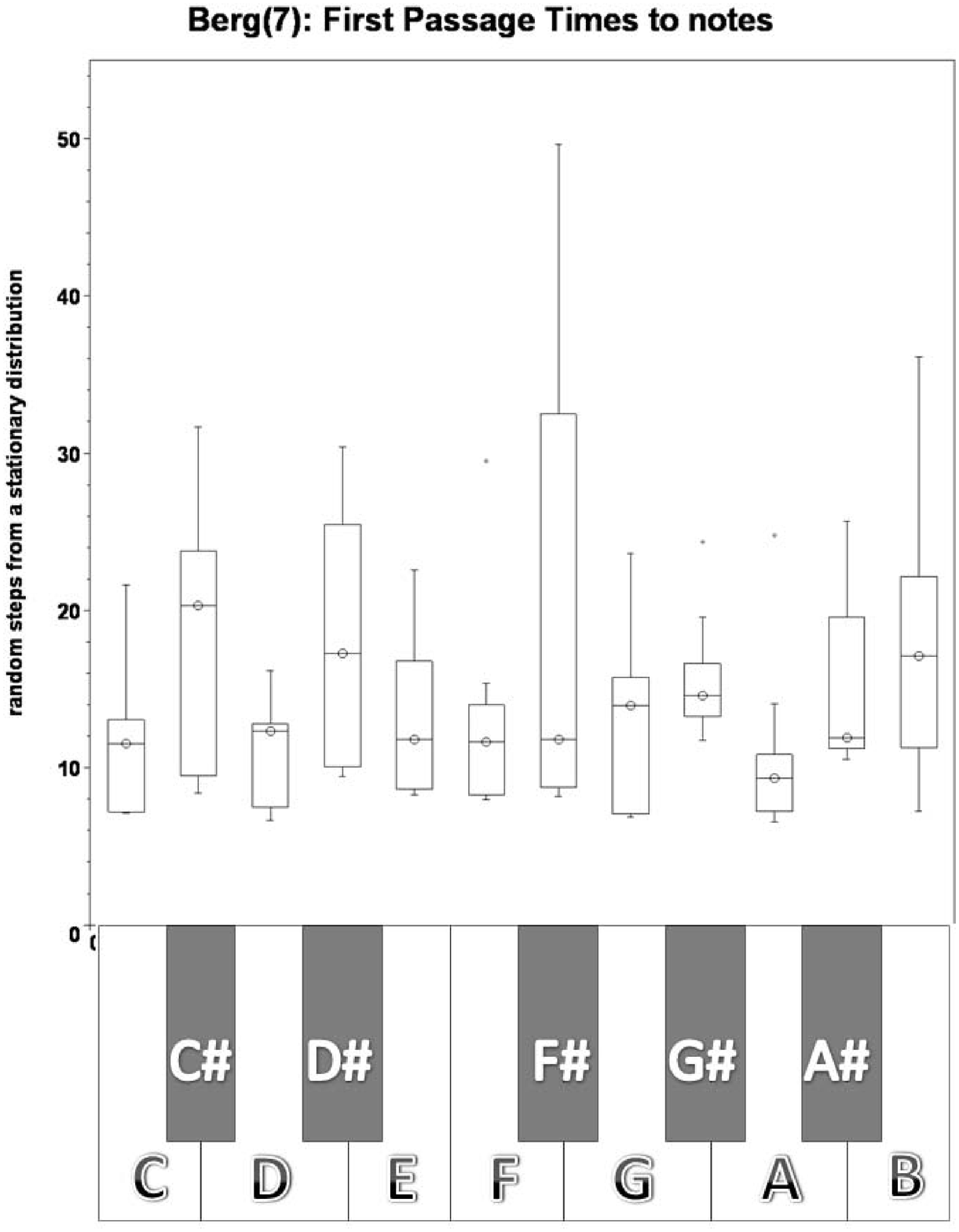, width=5cm, height =6cm} \\
\epsfig{file=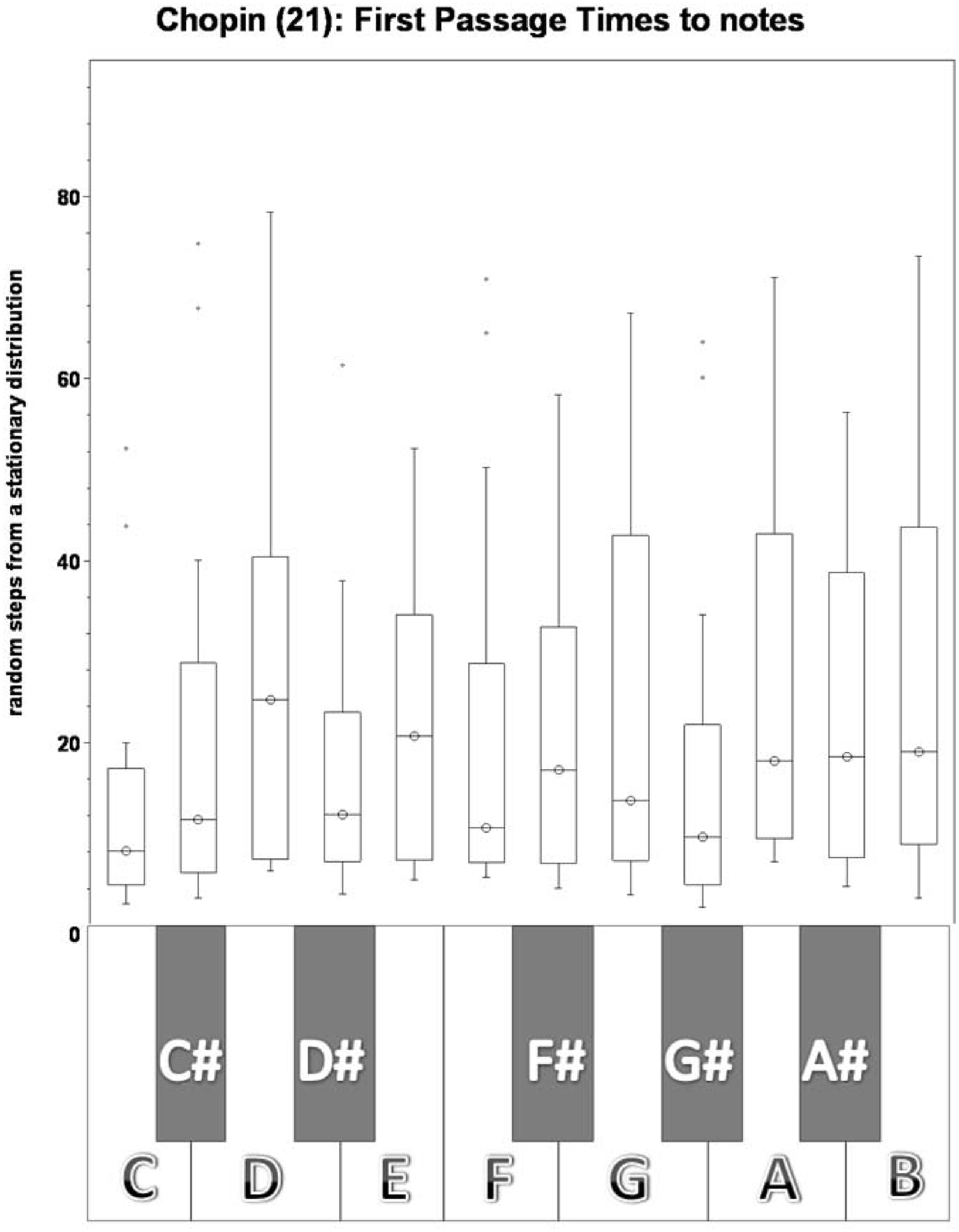, width=5cm, height =6cm}& 
 \epsfig{file=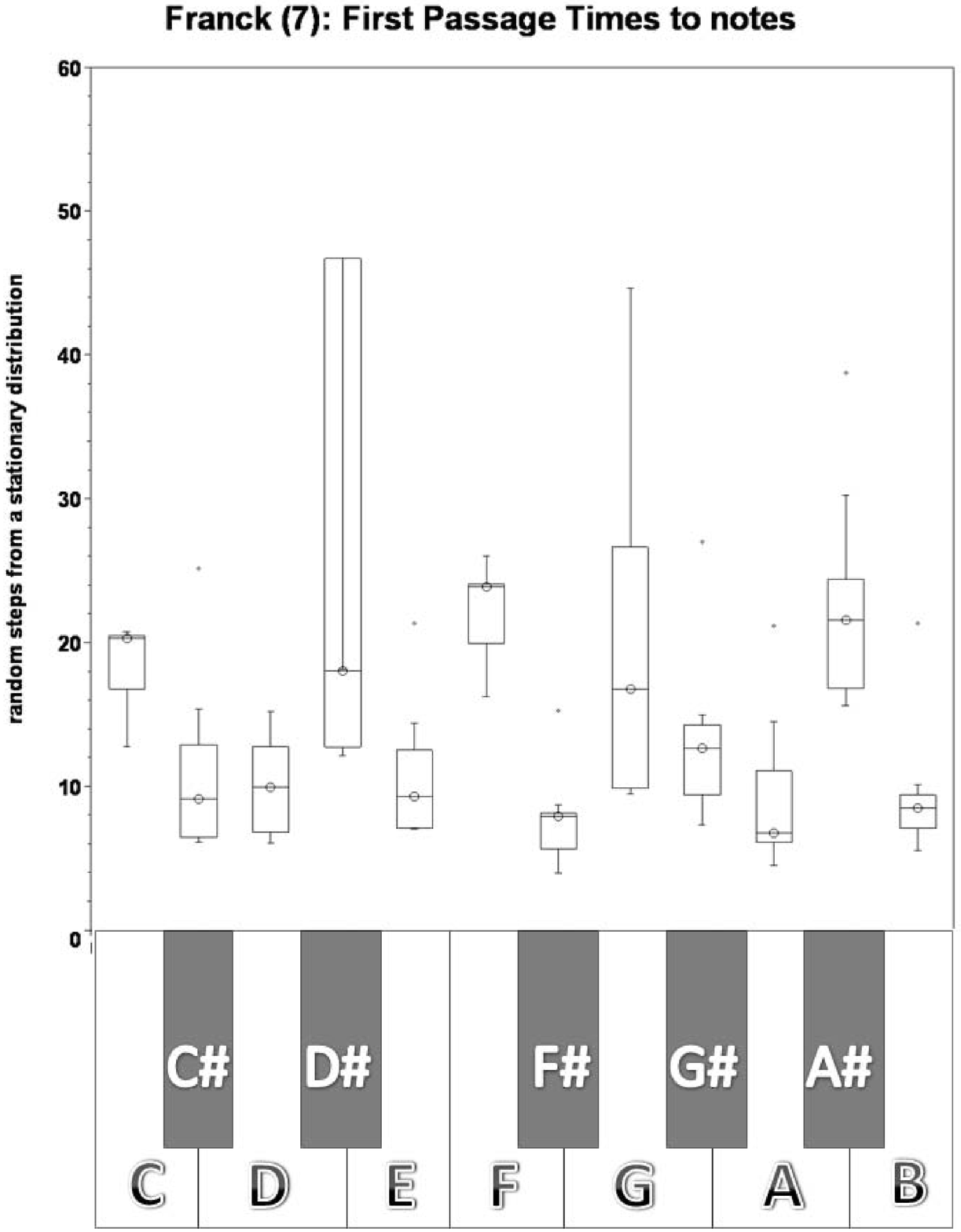, width=5cm, height =6cm}  
\end{tabular}
\end{center}
\caption{\footnotesize 
Statistics of first passage times of 
in the musical pieces 
of J.S. Bach, A. Berg, F. Chopin, and 
C.Franck are represented through their five-number 
summaries in the box plots.
 \label{Fig_music_07}}
\end{figure}
By analyzing the typical magnitudes
 of first passage times to notes in one octave,
we can discover
an individual music language of 
a composer and track out 
the stylistic influences between different composers. 
The box plots shown in 
 Fig.~\ref{Fig_music_07}
depict the data on first passage times to notes 
in a number of compositions written by 
J.S. Bach, A. Berg, F. Chopin, and 
C.Franck
through their five-number summaries: 
3/2 the  interquartile ranges,
the lower quartile, 
the third quartile,
and the median.
In tonal music, 
the magnitudes of first passage times to the notes
are completely determined by their roles in 
the hierarchy of tone scales.
Therefore, a low median in the box plot
(Fig.~\ref{Fig_music_07}) 
indicates that the note was often chosen  
as a tonic key in many compositions. 
\begin{figure}[ht!]
 \noindent
\begin{center}
\epsfig{file=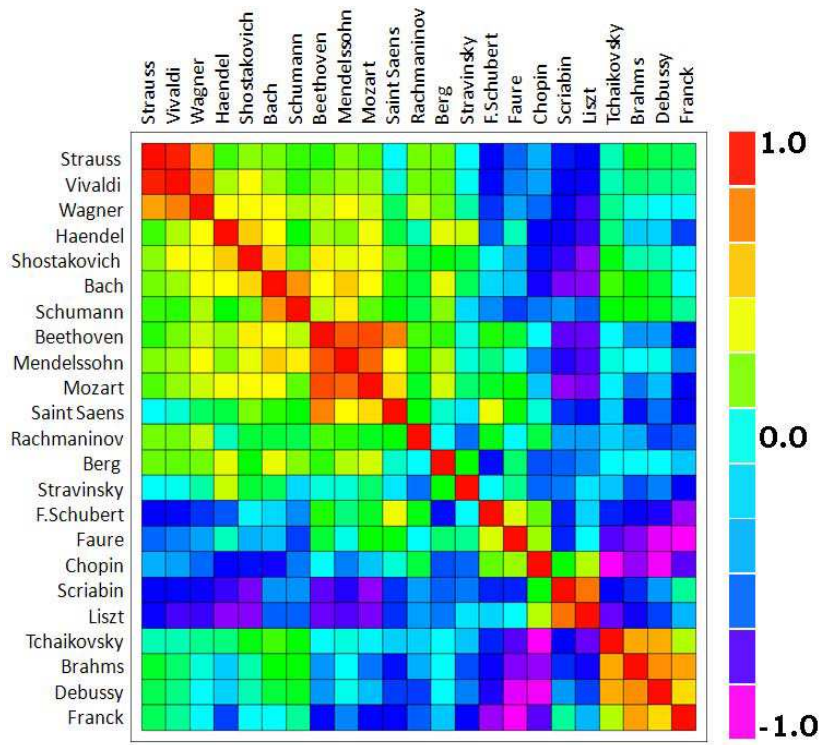,  angle= 0,width =11cm, height =10cm}
  \end{center}
\caption{\footnotesize The correlogram 
displays the correlation matrix
for the medians of the first passage times
to notes of one octave,
 for 23 composers. In the shaded rows, 
each cell is shaded from violet to red depending 
on the sign of the correlation, and with
the intensity of color scaled 0–-100\%, 
in proportion to the magnitude of the correlation.  \label{Fig_music_09}}
 \end{figure}
Correlation and covariance matrices
calculated for the medians of
 the  
 first passage times 
in a single octave
 provide 
 the basis 
for the classification of composers, 
with respect to their  
  tonality  preferences.
For our analysis, we have selected only
 those musical compositions,
in which all 12 pitches of the octave 
were used. 
The tone scale symmetrical correlation 
 matrix has been calculated for 23 composers,
 with the elements equal 
to the 
Pearson correlation coefficients 
between the medians of the first passage times.  
For exploratory visualization of the 
tone scale correlation matrix,
we arranged 
the "similar" composers
contiguously.
Following \cite{Friendly:2002}, 
while 
ordering the composers,
we considered 
 the eigenvectors (principal components)
of the correlation matrix
associated with its three largest eigenvalues.
Since the cosines of angles between 
the principal components 
approximate the correlations
 between the tonal preferences, 
we used an ordering based on the angular
 positions
of the three major eigenvectors 
to place the most similar composers contiguously, 
as it is shown 
in Fig.~\ref{Fig_music_09}.

 The correlogram  presented on Fig.~\ref{Fig_music_09}
allows for identifying the three groups of composers 
exhibiting similar preferences in the use of 
tone scales, as correlations are 
positive and strong  within each tone group while being 
weak or even negative between the different groups.
The smaller subgroups might be seen within 
the first largest group
(from J. Strauss to G. Faur\'{e}), in the left upper conner
of the matrix on Fig.~\ref{Fig_music_09}. 
Most of composers that appeared 
in the largest group  
are traditionally attributed to 
the Classical Period of music. 
The strongest  positive correlations 
we observed in the choice of a tonic key
(about 97\%)
 is between 
the compositions of J. Strauss and A. Vivaldi
who led  the way to a more individualistic assertion 
of imaginative music. The tonality statistics 
in the masterpieces of R. Wagner appears
also quite similar to them.
Other subgroups are formed by 
G.F. H\"{a}ndel and D. Shostakovitch,
J.S. Bach 
and R. Schumann. 
 The Classical Period boasted 
by L.V. Beethoven and W.A. Mozart who 
led the way further to the Romantic period
in classical music.
F. Mendelssohn Bartholdi 
 was deeply influenced by the music of
J.S. Bach, L.V. Beethoven, and 
W.A. Mozart, as often reflected by 
his biographers \cite{Brown:2003} --
not surprisingly, he
found his place next to them. 
Furthermore, the piano concerts of  
C. Saint-Sa\"{e}ns 
were  known to be strongly influenced
by those of W.A. Mozart,
 and, in turn,
 appear to have influenced those of 
 S. Rachmaninoff 
that receives full exposure 
in the correlogram  (Fig.~\ref{Fig_music_09}).
Moreover, 
we also get the evidence
of affinity between I.Stravinsky and A. Berg,
 F. Schubert, F. Chopin, and G. Faure,
as well as of
the strong correlation between
the tonality styles of   
A. Scriabin and F. Liszt.
The last group, in the lower right conner of the 
 matrix are occupied by the Middle and Late
Romantic era composers: P. Tchaikovsky, J. Brahms, C. Debussy, 
and C. Franck. 
Interestingly,
the names of composers that are 
contiguous in the correlogram  (Fig.~\ref{Fig_music_09})
are often 
found together in musical concerts and on records
performed by commercial musicians.

\section{On possible distances in space of musical dice games}
\label{sec:musical_distances}

Most music is written for playing on standard keyboards 
and involves
 mostly overlapping sets of pitches. 
Given two pieces modeled 
by the different musical dice games 
but defined on the same set of pitches,
a natural idea arises to compare their Markov chains 
in order to estimate their similitude. 

Let us note that 
the  Kullback –- Leibler divergence \cite{Cover:1991},
a measure of the difference
between two probability distributions
playing 
the important role in information theory,
cannot help us much 
with the Markov transition matrices 
since 
the transition probability vectors 
(rows of the transition matrices)
in general
are not the probability distributions, as 
many of their components
 might be equal to zero
(even for quite a long composition mapped 
into one octave)
 thus prohibiting 
transitions between some states of the 
Markov chain. 
Nevertheless, 
the  Kullback –- Leibler divergence
can be used in order to compare 
two different musical dice games
defined on the same set of pitches
by means of their 
stationary vectors
(\ref{music_stationary_vector}),
\begin{equation}
\label{musical_Kullback_leibler}
D_{KL}\left(\pi^{(1)}\mid\pi^{(2)}\right)\,\,=\,\,
\sum_{i=1}^n \pi^{(1)}_i\log\left(\frac{\pi^{(1)}_i}{\pi^{(2)}_i}\right).
\end{equation}
 The Kullback–Leibler divergence
(\ref{musical_Kullback_leibler})
 is neither symmetric,
nor satisfies the triangle inequality.

The Euclidean distance
 between 
  the two transition 
matrices, 
${\bf T}_A$ and ${\bf T}_B,$ 
is defined by
\begin{equation}
\label{music_Euclidean_distance}
\mathcal{D}^{(E)}_{AB}\,\,=\,\,\left\|
{\bf T}_A -{\bf T}_B
 \right\|_F
\end{equation} 
where $\|\ldots\|$ is the Frobenius norm 
 induced by the Euclidean inner product
for matrices,
$
\left({\bf T}_A,{\bf T}_B\right)\,\,=\,\,\mathrm{Tr}
\left({\bf T}_A^\top{\bf T}_B\right),
$
in which ${\bf T}^\top$ denotes a transposed matrix.
However, there is 
no any indication of that 
 probabilistic space of 
musical dice games  
possesses the structure of Euclidean space. 

Another possibility 
to compare the musical 
dice games by their transition 
matrices is to 
use the Riemann structure associated to 
the probability vectors (rows of the transition matrices) 
instead of (\ref{music_Euclidean_distance}).
Let us discuss such a distance in more details, 
for the case of $12\times 12$ transition matrices
$\bf T$. 

First,  let us introduce the new matrix $Q_{ij}=\sqrt{T_{ij}}$
and note that the 12 rows of ${\bf Q}$ define 
the 12 points 
on the surface of a unit sphere $S_1^{11}.$
It is obvious that 
under any change to the musical dice game
the 
rows of the matrix 
${\bf Q}$ remain on the surface of $S_1^{11},$
and therefore 
the difference between 
a pair of musical compositions 
chromatically transposed into one octave
is  always described by a set of 12 
rotations 
$\left\{\omega_1,\ldots,\omega_{12}\right\}\in \mathrm{SO}(12)$
relating the two sets of 12 points on $S_1^{11}.$

Second, given ${\bf Q}_A$ and ${\bf Q}_B,$ 
representing the two different musical dice games,
$A$ and $B,$ on 
$S_1^{11}$,
 we can approximate the set of rotations
$\left\{\omega_1,\ldots,\omega_{12}\right\}$
by  a single one 
$\Omega_{AB}\in \mathrm{SO}(12)$
that minimizes the Frobenius norm of
 a possible  discrepancy,
$$
\min_{\Omega\in\mathrm{SO}(12)}\left\|
{\bf Q}_A\Omega_{AB}-{\bf Q}_B
\right\|_F.
$$
Indeed, such a minimization
is nothing else but 
the orthogonal Procrustes problem \cite{Gower:2004},
which is equivalent to 
the singular value decomposition
of the matrix ${\bf Q}_A^\top{\bf Q}_B$,
\begin{equation}
\label{music_SVD}
{\bf Q}_A^\top{\bf Q}_B\,\,=\,\,
{\bf U}\Sigma{\bf V}^\top,\quad 
\Omega_{AB}\,\,=\,\,{\bf U}
{\bf V}^\top.
\end{equation}
The matrix $\Omega_{AB}\in\mathrm{SO}(12)$
defined in (\ref{music_SVD}) describes 
the optimal rotation 
(with respect to the Frobenius norm) 
 translating ${\bf Q}_A$ to ${\bf Q}_B,$
while the transposed matrix, $\Omega_{AB}^\top,$
makes the backward translation, ${\bf Q}_B$
to  ${\bf Q}_A$. Obviously, 
$\Omega_{AB} ={\bf 1}$ if and only if ${\bf Q}_A={\bf Q}_B.$

We define the Riemann distance between 
the two musical dice games, $A$ and $B$,
as the length of a geodesic curve 
connecting ${\bf Q}_A$ and ${\bf Q}_B$
on the surface of $S_1^{11}$
\begin{equation}
\label{music_geodesic_distance}
\mathcal{D}^{(R)}_{AB}\,\,=\,\,
\left\|\log\,\, \Omega_{AB} \right\|_F.
\end{equation}
It follows from the definition
that  
the metric (\ref{music_geodesic_distance}) 
satisfies the conditions of 
non-negativity, identity of indiscernibles,
symmetry, and subadditivity. The 
triangle inequality is satisfied, 
as the length of the geodesic curve 
on the unit sphere,
$\exp\left(t\,\,\log\,\Omega_{AB}\right)$,
$0\leq t \leq 1,$ is a strictly positive 
 function.

\section{Conclusions}
\label{sec:music_conclusions}

We have studied the musical dice games 
encoded by the transition matrices 
between pitches 
of the 804 musical compositions. 
Contrary to the language
where the alphabet is independent of a 
message, musical compositions 
might involve different sets of pitches;
the number of pitches used to compose a piece 
grows approximately logarithmically with its size. 

Entropy dominates over redundancy in the musical dice games
derived from the compositions of classical music. 
Thus the successful understanding of a musical 
composition requires much attention and experience 
from a listener. Statistics of complexity 
measured by the past-future mutual information 
suggests that pieces in classical music 
might contain a few melodic lines
translated over the diapason of pitches by chromatic 
transposition. 
The hierarchical relations between pitches in tonal  
music can be rendered by means of first passage time 
to them, in musical dice games. Correlations between the 
medians of the first passage times to the notes of one octave
provide the basis for the classification of composers, 
with respect to their tonality preferences. 
Finally, we have discussed 
the possible distances in space of musical 
dice games and introduced the geodesic
distance based on the
 Riemann structure associated to the probability vectors
 (rows of the transition matrices).

\section{Acknowledgments}
\label{sec:Acknowledgments}
We thank Ph. Blanchard, T. Kr\"{u}ger, and J. Loviscach
for the inspiring discussions.


\begin{thebibliography}{000}

 
\bibitem{Noguchi:1996}
H. Noguchi, "Mozart: Musical game in C K.516f". Available at {\it http://www.asahi-net.or.jp/~rb5h-ngc/e/k516f.htm} (1996).

\bibitem{Markov:1906}
A.A. Markov. "Extension of the limit theorems of probability theory to a sum of variables connected in a chain". reprinted in Appendix B of: R. Howard. 
{\it Dynamic Probabilistic Systems}, vol. {\bf 1}: Markov Chains. 
John Wiley and Sons (1971). 

\bibitem{Shannon:1948}
C.E. Shannon, "A mathematical theory of communication," 
{\it Bell System Technical Journal} {\bf 27}, pp. 379-423; 623-656 
(1948). 
\bibitem{Shannon:1951}
C.E. Shannon, 
"Prediction and entropy of printed English",
{\it  The Bell System Technical Journal} {\bf 30}, p. 50 (1951). 
\bibitem{Wolfe:2002}
J. Wolfe, "Speech and music, acoustics and coding,
and what music  might be 'for'."
{\it Proc. the 7th International Conference on Music 
Perception and Cognition}, Sydney, 2002;
C. Stevens, D. Burnham, G. McPherson, E. Schubert, 
J. Renwick (Eds.).  Adelaide: Causal Productions.
\bibitem{Seeger:1971}
Ch. Seeger, "Reflections upon a Given Topic: 
Music in Universal Perspective". 
{\it Ethnomusicology} {\bf 15}(3), p. 385 (1971).
\bibitem{Mutopia}
All music in the Mutopia Project
 free to download, print out, perform and distribute
is available at {\it http://www.mutopiaproject.org}.
While collecting the data, 
we have also used the following free resources:
{\it http://windy.vis.ne.jp/art/englib/berg.htm} 
(for Alban Berg), {\it http://www.classicalmidi.co.uk/page7.htm},
{\it http://www.jacksirulnikoff.com/}.

\bibitem{Perl_MIDI}
This software is freely available at {\it 
http://search.cpan.org/~sburke/MIDI-Perl-0.8.}

\bibitem{Schurmann:1996}
T. Sch\"{u}rmann, P. Grassberger, 
"Entropy Estimation of Symbol Sequences",
 {\it CHAOS} {\bf 6}(3)  414-427 (1996).
\bibitem{Cover:1991}
T.M. Cover,  J.A. Thomas,   
{\it Elements of Information Theory}. London: Wiley (1991).
\bibitem{Meyer:1975} 
C.D. Meyer, "The role of the group generalized inverse in the theory of finite Markov chains", {\it SIAM Rev.} {\bf 17}, p. 443 (1975).


\bibitem{Meyer:1982}
C.D. Meyer, "Analysis of finite Markov chains
 by group inversion techniques.
 Recent Applications of Generalized Inverses", 
in: S.L. Campbell (Ed.), 
{\it Research Notes in Mathematics} {\bf 66}, 
Pitman, Boston, p. 50 (1982).
\bibitem{Drazin:1958}
M.P. Drazin,  
"Pseudo-inverses in associative rings and semigroups".
 {\it The American Mathematical Monthly} {\bf 65}, 506-514 
(1958).

\bibitem{Ben-Israel:2003}
 A. Ben-Israel, Th.N.E. Greville, {\it Generalized inverses: theory and applications.}
Springer; 2nd edition (2003).

\bibitem{Erdelyi:1967}
I. Erd\'{e}lyi, "On the matrix equation $Ax=\lambda Bx$."
{\it J. Math. Anal. Appl.} {\bf 17}, 119-132 (1967).
\bibitem{Campbell:1979}
S.L. Campbell, C.D. Meyer, {\it Generalized Inverses 
of Linear transformations.} New
York: Dover Publications (1979).

\bibitem{Lin:1973}
 N. Lin,  {\it The Study of Human Communication},
 The Bobbs-Merrill Company, Indianapolis, (1973).
\bibitem{Shaw:1984}
R. Shaw, {\it The dripping faucet as a model chaotic system},
CA Aerial Press, Santa Cruz (1984).

\bibitem{Li:1991}
W. Li, "On the relationship Between Complexity 
and Entropy for Markov Chains and Regular Languages",
{\it Complex systems} {\bf 5}, 381 (1991).

\bibitem{Thomson:1999}
W. Thomson, {\it Tonality in Music: A General Theory.},
 San Marino, Calif.: Everett Books  (1999).
\bibitem{Blanchard:2008}
Ph. Blanchard, D. Volchenkov, 
"Intelligibility and first passage times in 
complex urban networks", {\it Proc. R. Soc. A} {\bf 464}, 2153 (2008).
\bibitem{Volchenkov:2010}
D. Voclchenkov, "Random walks and flights over connected graphs and
 complex networks", {\it Commun. Nonlinear Sci. Numer. Simulat},
doi:10.1016/j.cnsns.2010.02.016 (2010).
\bibitem{Burns:1999}
E.M. Burns, {\it Intervals, Scales, and Tuning.
 The Psychology of Music}.
 Second edition, Deutsch, Diana, ed. San Diego: Academic Press (1999). 

\bibitem{Lovasz:1993}
 Lov\'{a}sz, L. 1993 Random Walks On Graphs: A Survey. {\it Bolyai Society Mathematical
 Studies} {\bf 2}: {\it Combinatorics,
Paul Erd\"{o}s is Eighty}, Keszthely (Hungary), p. 1-46.

\bibitem{Friendly:2002}
M. Friendly, "Corrgrams: Exploratory Displays for Correlation Matrices". 
{\it The American Statistician} {\bf 56}(4), 316
 (2002). 

\bibitem{Brown:2003}
C. Brown,  {\it A Portrait of Mendelssohn},
 New Haven and London (2003). 


\bibitem{Gower:2004}
 J.C. Gower, G.B. Dijksterhuis, 
{\it  Procrustes Problems}. Oxford University Press   (2004).












\end{thebibliography}
\end{document}